# The Effect of Whole-Body Haptic Feedback on Driver's Perception in Negotiating a Curve

Erfan Pakdamanian[1], Lu Feng[1,2], Inki Kim[1]
[1] Systems and Information Engineering, University of Virginia, Charlottesville, VA
[2] Computer Science, University of Virginia, Charlottesville, VA

It remains uncertain regarding the safety of driving in autonomous vehicles that, after a long, passive control and inattention to the driving situation, how the drivers will be effectively informed to take-over the control in emergency. In particular, the active role of vehicle force feedback on the driver's risk perception on curves has not been fully explored. To investigate it, the current paper examined the driver's cognitive and visual responses to the whole-body haptic feedback during curve negotiations. The effects of force feedback on drivers' responses on curves were investigated in a high-fidelity driving simulator while measuring EEG and visual gaze over ten participants. The preliminary analyses of the first two participants revealed that pupil diameter and fixation time on the curves were significantly longer when the driver received whole-body feedback, compared to none. The findings suggest that whole-body feedback can be used as an effective "advance notification" of hazards.

## INTRODUCTION

Although the future of car industries will be dominated by autonomous vehicles and the car will drive itself, there will still be a need for drivers to take over the car (Banks & Stanton, 2016). Human intervention is particularly necessary to prevent tragic accidents when the autonomous vehicle encounters curves, bad weather and unpredictable pedestrian behavior (Wright, Svancara, & Horrey, 2017). Although autonomous vehicles will overall decrease the physical and mental workload of drivers by assigning these tasks to an automated system, human drivers would still play a critical role in car safety responsibility (Parasuraman and Wickens, 2008). However, it was been shown that a sudden alarm and notification to the driver about the upcoming potential hazards would incur higher stress and cognitive load (Shah et al., 2015). Once drivers allow the automated system to control the car, meaning the driver tends to allocate his attention resources to non-driving tasks (e.g. video gaming, talking on the phone etc.), his or her attention will be taken away from the primary task of driving. In such circumstances, any simple form of visual, auditory or haptic signals would not be sufficient to communicate critical information about the vehicle conditions, only to startle and stress the driver in emergency (Petermeijer, Cieler, & Winter, 2017).

One approach to cope with the stress from unexpected alarms is to examine the effects of signals on potentially safety-compromising situations, and accustom the driver to it. In this regard, this paper intends to investigate the effects of whole-body haptic feedback, delivering haptic cues to drivers' full body, on the drivers' visual perception and cognitive states during curve negotiation, as an alternative to its counterpart alarming signals. Assessing the drivers' cognitive states can help infer what type of haptic feedback the cars should provide to mitigate the stress of taking over during critical moments. In literature, vibrotactile haptic feedback was shown to enhance the reaction time of taking control back at life-threatening moments (Prewett, Elliott, Walvoord, & Coovert, 2012). It noted, however, that once the drivers were spatially aware, the vibrotactile "directional" cue may not be as effective as visual directional alternatives. Therefore, this study intends to focus on whole-body haptic feedback to complement this drawback.

Morrell and Wasilewski (2010) designed and developed a haptic-feedback seat for traditional vehicles that aimed to share spatial information, and improve situation awareness (SA). The drivers were informed about the location of car-following and close-by vehicles, through vibrotactile feedback from the seat back in a way that the closer the car is, the more sensors vibrated. Nonetheless, on the one hand, evaluating the time in the blind spot may not be the accurate measurement for the risk assessment. Nonetheless, on the one hand, evaluating the time in the blind spot may not be the accurate measurement for the risk assessment. On the other hand, as auto industries attend to autonomous technologies, alert systems need to become adaptive to vehicle speed and situation but not particularly designed for a specific scenario.

Petermeijer et al. (2017) designed a vibrotactile feedback seat that contains static and dynamic vibration for automated vehicles. The authors aimed to analyze the accuracy of drivers' response rate and their reaction time to the requested time for maneuvering. After receiving tactile stimuli, drivers had to respond accordingly to the vibration direction by moving to the left or the right. However, in all the presented scenarios, there was not any additional warning cue. Furthermore, the participants reported difficulties in understanding whether the cue was to their left or right; alarms were only triggered about one second prior to an event occurrence, which was shorter than the realistic average reaction time needed (3.5 sec) for a transition control in automated vehicles (Melcher, Rauh, Diederichs, Widlroither, & Bauer, 2015).

This research aims to examine perceptual and cognitive effects of using whole-body force feedback on the control responses of the drivers. Through the controlled experiments in simulation setting, it is expected that the whole-body force feedback will be shown its values, in a way that does not only warn the driver when a takeover is required, but also assists the driver during the critical phases, including their lack of SA (shifting of attention) and cognitive processing. In this regard, we hypothesized that the whole-body haptic feedback would allow the drivers to be effectively aware of upcoming curves in a simulated driving environment.

## METHODS

The experiment was conducted in a high-fidelity driving simulator (the 401cr motion system by Force Dynamics) equipped with three monitors. The simulator mimics various acceleration dynamics thereby creating a realistic response upon the driver's body. The motion-capable high-fidelity simulator was used with two configurations: 1) without whole-body motion feedback, 2) with whole-body motion feedback with approximately 18 inches of movement in 360 degrees. This also allowed six degrees of freedom to replicate the motions associated with driving in a way that vibration of the seat serves as an "intelligent messenger". It ensures human stays informed of the vehicle safety. The study was approved by University of Virginia Institutional Review Board (Protocol # 2017-0296-00).

The speedometer and the RPM gauge is located in the center of the middle monitor (Figure. 4). Moreover, the implemented automation system had a longitudinal capability similar to common ACC systems, which allow drivers to follow the indicated speed limits as well as keep the car in the center of the lane. Data were recorded at a frequency of 100 Hz, including the vehicle's position, accelerations and steering wheel angle (they were not included in the preliminary study and will be reported in further analysis).

### Data acquisition

A wearable eye-tracker glass (Tobii Pro-Glasses 2, Danderyd, Sweden; Tobii Pro-Glasses 2, 2017) was used to track the driver's gaze behavior at a sampling rate of 60 Hz (i.e., 60 gaze data points collected per second for each eye; 4 eye cameras, H.264 1920x1080 pixels at 25 fps) (Figure 1). The Tobii Pro Glasses 2 eye-tracker is wireless with live view capability for insights in any real-world environment. Since the driving simulator and curves are dynamic scenes, head-mounted eye tracker was required. Also, it ensures that the participant's full and complete range of motions for their head.

A B-Alert X24 system with 24 channels was used with the sample rate of 256Hz to record the Electroencephalography (EEG) data (Figure 1). Wireless EEG signals were sent via Bluetooth to the data acquisition system. Also, in order to record the electoral activity of the brain, the sensor strip was placed according to the 10/20 extended standard.

The sampled data was sent wirelessly to iMotion (biometric research platform) which allowed collection of the synchronized EEG and eye-tracker data (Attention tools, 2016).

### Procedure

Two graduate students (both male, 22 and 35 years old) holding a driver's license voluntarily participated in this preliminary study (ten participants equally balanced between male and female aged between 18 to 40 will be recruited). None of the participants had visual impairments, or any other symptoms or diseases that could compromise their ability to drive.

Once the participant arrived, the relevant information regarding gender, age and driving experiences was gathered. Subsequently, participants were verbally instructed regarding how to use the devices and simulator as well as their primary task of driving with their hands on the wheel by the experimenter. Furthermore, both drivers were told that they need to keep their speed under 60mph and drive as they would normally do. The experimenter allowed the participants to familiarize themselves with the system with 2-5 mins test drive. Once they showed that they were comfortable with all the devices and driving the simulator, there were asked to take 3mins break between the sessions in order to maximize the concentration level and minimize fatigue throughout the 18 min session. The experimenter started the three curve and force-feedback-free trials as the Baseline session. Afterwards, the participants drove through the counterbalanced designated scenario six times (three trials with force feedback and three without). Each scenario took approximately 3mins, depends on the speed.

**Table 1. Mean and Standard Deviation for Metrics of Eye Movements**

| Dependent Variables | Independent Variables | | |
|---|---|---|---|
| | With force feedback | Without force feedback | p-value |
| Time spent (sec) | 6.83 (1.92) | 4.49 (0.28) | 0.002 |
| Fixation duration (sec) | 3.45 (0.89) | 3.04 (0.61) | <0.001 |
| Pupil diameter (px) | 35 (9.5) | 27 (4.3) | 0.008 |

### Signal Pre-processing

256Hz sampled data was filtered using high and low band pass filter with a cut-off frequency between 0.5 Hz, to remove DC drift, and 80 Hz respectively to remove power-line noise and low frequencies separately (Gheorghe, 2017). Also, a notch filter at 60Hz was used. EEG data pre-processing initiated by referencing to the left ear lobe channel as well as applying Fast Fourier transform (FFT) algorithm to filter the different frequency band.

To analyze the EEG data, initially the blink artifacts were removed by using Independent Component Analysis (ICA) and wavelet analyses were used to generate a continuous record of theta band by using Matlab (2017, The MathWorks, Inc., Natick, Massachusetts, United Statest) and EEGLab toolbox. An electrode impedance test was performed to ensure proper conductivity of the electrodes. The impedance level threshold of 20 kΩ was used. Also, the EEG calibration procedure was implemented before data collection.

### RESULTS

Collected data were extracted using iMotion software. In order to perform a comparison analysis between three conditions (Baseline, with whole body force feedback and without), approximately four seconds before curves was analyzed following the approach taken by Gheorghe (2017). Each trial consisted of twenty curves, including simple curves, compound curves, reverse curves and deviation. However, we were only interested in simple curves for our preliminary study.

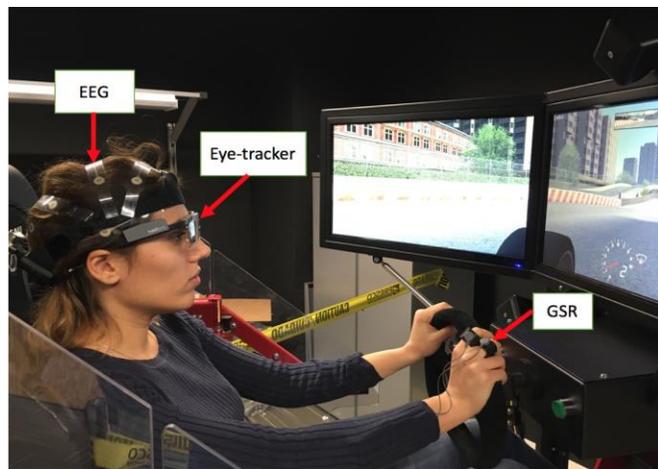

**Figure 1. Experimental set-up for recording EEG and Eye movement**

### Analysis of visual attention

iMotion provides the following metric for analyzing eye movement: Time spent-fixation, fixation duration, Time to First Fixation (TTFF-F) and pupil diameter (Table 1). Table 1 summarizes the time spent and fixation duration on the AOI. In order to identify when curves as the critical section of the road on the visual display were fixed, AOI analysis was performed (Figure 4). Comparing the TTFF values (Figure 3) indicates both participants tended to concentrate slightly more on the curves at the presence of force feedback which indicates higher SA. Likewise, when the force feedback was applied to drivers, pupil diameter was larger approaching the curves (Table 1). Therefore, the drivers tend to fixed their gaze on curves significantly higher at the presence of the whole-body feedback.

The differences between the two types of vibration patterns including force and none-force feedback was assessed using t-test. T-test yielded statistically difference between the force feedback and none in the dependent variables (t(11) = 4.96, p= 0.002; t(11) =12.38, p<0.001; t(11) = 3.51, p=0.008, for time spent, fixation duration, and pupil diameter, respectively).

### Analysis of cognitive states

Analyzing three frequencies (Theta, Alpha and Beta) revealed that the Theta power increases in force feedback cases. Also, on the beta band, grown power was obtained. Still, the amount of power increasing on Theta band was higher, which may indicate the greater drivers' engagement while using haptic feedback. The findings represent that the force-feedback could correlate with higher ability in decision making and ultimately increase the capability of controlling the vehicle properly at the time of hazard encounter. It was initially expected to get the consistent results with Almahasneh et al. (2014) findings, however, the topographical map result (Figure 3) indicates that the difference between baseline and both cases is caused by more activity in corresponding brain region of the right frontal hemisphere near reaching the curves. Since most of cognitive activities occur at the frontal lobe (Lin et al., 2011) the findings are aligned with the role of frontal lobe in decision making and attention (Burgess, Alderman, Volle, Benoit, & Gilbert, 2009). The topographical analysis extracted from the scalp above the sensorymotor cortex indicates more activity on the bipolar channels C3 and C4 (Figure 3). Electrode C4 represents the highest activation throughout the six curves which may cause by Motor execution phase of driving. Slightly higher activation in motor cortex at the presence of whole-body haptic feedback supports an enhancement to drivers' engagement of required cognitive tasks of braking and steering control (Saha, Konar, Nagar, 2017). However, band frequency modulation based on ERP will be analyzed at the critical time intervals of curve negotiation. Our intent is to analyze the variability of frequency bands inside some temporal windows around 200 ms and 400 ms of latency.

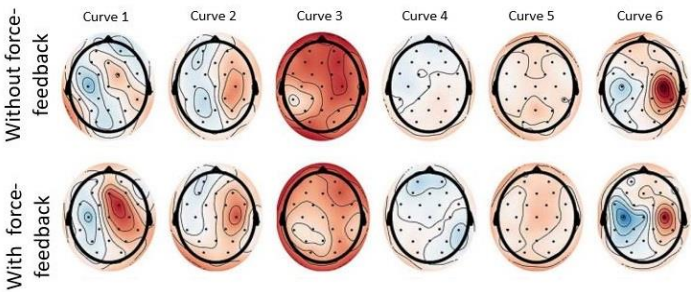

**Figure 2. Topographical analysis of six simple curves. The first row represents the distribution of difference between baseline and scenarios at the absence of force feedback (the first row) and the at the presence of it (the second row)**

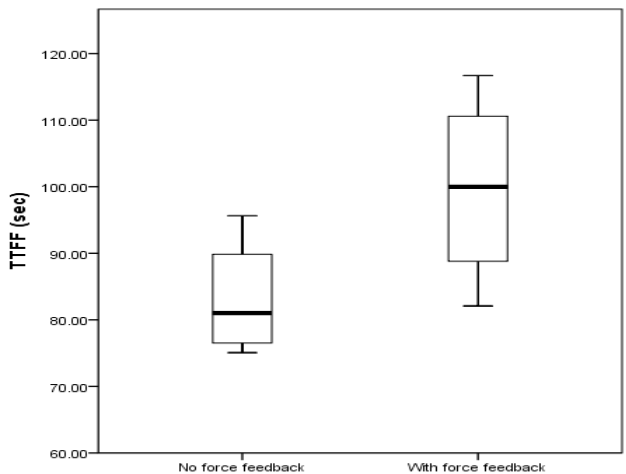

**Figure 3. Eye- tracking results for fixation behaviors over different feedback condition**

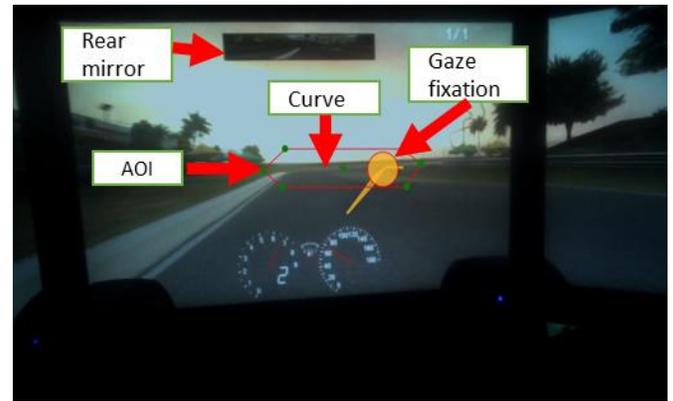

**Figure 4. AOI and driver's view**

## DISCUSSION

The main differences between the two types of feedback found in this study is containing driver's visual responses. Fixed duration and pupil diameters found significantly higher while driving with haptic feedback in this preliminary study which could be due to the higher cognitive engagement. If it was the case, finding higher power in Theta band in frontal lobe is due to high vibration of system during haptic feedback activation and it is not relevant to the type of feedback. Therefore, the findings could be supported by the results that the high-fidelity driving simulator that can simulate various scenarios with high validity improvement of drivers' performance engages driver better (Groeger & Banks, 2007).

This preliminary study confirms the possibility of EEG usage to alarm drivers properly within less than few seconds, once the system recognizes driver's cognitive stage and driving environment. We expect that the need for more number of channels for prediction of performance and drivers' cognitive state prior to hazard with other EEG measurements (e.g. ERP) would help us to develop a safer whole-body feedback to reduce cognitive workload and stress level of the driver, thereby enhance their control ability.

In the future, we will design and analyze a haptic force feedback which could communicate with drivers through the seat and serves as an "intelligent messenger" that ensures human stays informed of the vehicle safety as well as driving environment which could play the role of "advance notification". In that regards, we will validate the preliminary findings with further analysis of power variation in each frequency within temporal duration as well as Event Related Potential (ERP). It could assist us to identify the perceptual operations of drivers on curves.